\documentclass[conference]{IEEEtran}
\usepackage[table]{xcolor}
\usepackage{booktabs}
\usepackage{multirow}
\usepackage{color}
\usepackage{subcaption}
\usepackage{algorithm}
\usepackage{algorithmicx}
\usepackage{algpseudocode}
\usepackage{multirow}

\usepackage{cite}

\ifCLASSINFOpdf
   \usepackage[pdftex]{graphicx}
\else
\fi
\begin{document}
%
\title{PeerHunter: Detecting Peer-to-Peer Botnets through Community Behavior Analysis \vspace{-0.2cm}}


%
\vspace{-1.12cm}
\author{\IEEEauthorblockN{Di Zhuang\IEEEauthorrefmark{1},
J. Morris Chang\IEEEauthorrefmark{2}}
\IEEEauthorblockA{Department of Electrical Engineering,
University of South Florida, Tampa, Florida 33620\\
Email: \IEEEauthorrefmark{1}dizhuang@mail.usf.edu,
\IEEEauthorrefmark{2}chang5@usf.edu} \vspace{-1.0cm}}

\IEEEoverridecommandlockouts
\IEEEpubid{\makebox[\columnwidth]{978-1-5090-5569-2/17/\$31.00~
\copyright2017
IEEE \hfill} \hspace{\columnsep}\makebox[\columnwidth]{ }}

\maketitle
\begin{abstract}
Peer-to-peer (P2P) botnets have become one of the major threats in network security for serving as the infrastructure that responsible for various of cyber-crimes. Though a few existing work claimed to detect traditional botnets effectively, the problem of detecting P2P botnets involves more challenges.
In this paper, we present PeerHunter, a community behavior analysis based method, which is capable of detecting botnets that communicate via a P2P structure. PeerHunter starts from a P2P hosts detection component. Then, it uses mutual contacts as the main feature to cluster bots into communities. Finally, it uses community behavior analysis to detect potential botnet communities and further identify bot candidates. Through extensive experiments with real and simulated network traces, PeerHunter can achieve very high detection rate and low false positives.
\end{abstract}


%
\IEEEpeerreviewmaketitle

\setlength{\textfloatsep}{0.6pt}
\section{Introduction}
\label{sec:sec1}
A botnet is a set of compromised machines controlled by botmaster through a command and control (C\&C) channel. 
Botnets may have different communication architectures.
Classical botnets were known to use a centralized architecture, which has a single point of failure. Peer-to-peer (P2P) network happens to be modeled as a
distributed architecture, where even though a certain number of peers fail to function properly, the whole network is not compromised. In this case, the most of recent botnets (e.g. Storm, Waledac, ZeroAccess, Sality and Kelihos) attempt to build on P2P network, and P2P botnets have proven to be highly resilient even after a certain number of bots being identified or taken-down \cite{rossow2013sok}. P2P botnets provide a fundamental infrastructure for various cyber-crimes \cite{wang2010peer}, such as distributed denial-of-service (DDoS), email spam, click fraud, etc. Therefore, detecting P2P botnets effectively is rather important for cyber security.

However, designing an effective P2P botnets detection system is extremely hard, due to several challenges.
First, botnets tend to act stealthily \cite{zhang2014building} and spend most of their time in the waiting stage before actually performing any malicious activities \cite{wang2009botnet, hang2013entelecheia}. Second, botnets tend to encrypt C\&C channels, which makes deep-packet-inspection (DPI) based methods fail to work. Third, botnets can randomize their communication patterns
dynamically without jeopardizing any primary functions \cite{stinson2008towards, gu2008botminer, zhang2011detecting}, which makes statistical traffic signatures based methods unable to work.

In this paper, we propose PeerHunter, a novel community behavior analysis based P2P botnet detection system, which could address all the challenges above. We consider a botnet community as a group of compromised machines that communicate with each other or connect to the botmaster through the same C\&C channel, are controlled by the same attacker, and aim to perform similar malicious activities. Due to the dynamic changes of communication behaviors of P2P botnets \cite{yan2013peerclean}, it would be extremely hard to identify a single bot. However, bots within the same P2P botnet always work together as a community, thus, have distinct community behaviors to be identified. PeerHunter begins with a general P2P hosts detection component.
Then, it builds a mutual contact graph (MCG) of the detected P2P hosts. Afterwards, it applies a community detection method on the MCG, which uses mutual contacts \cite{coskun2010friends} as the main feature of P2P botnets to cluster bots within the same botnet together, and separate bots and legitimate hosts or different types of bots into different communities.
Finally, it uses $destination \ diversity$ and $mutual \ contacts$ as the natural features to capture the ``P2P behavior'' and ``botnet behavior'' respectively of each P2P botnet community, and further identify all the P2P botnets.

Specifically, PeerHunter is capable of detecting P2P bots with the following challenges and assumptions: (a) botnets are in their waiting stage, which means there is no clear malicious activity can be observed \cite{wang2009botnet}; (b) the C\&C channel has been encrypted, so that no deep-packet-inspection (DPI) can be deployed;
(c) no bot-blacklist or ``seeds'' information \cite{coskun2010friends} are available; (d) none statistical traffic patterns \cite{yan2013peerclean} known in advance; and (e) could be deployed at network
boundary (e.g. gateway), thus, do not require to monitor individual host.

In the experiments, we mixed a real network dataset from a public traffic archive
\cite{mawi} with several P2P botnet datasets and legitimate P2P network datasets \cite{rahbarinia2014peerrush}.
To make the experimental evaluation
as unbiased and challenging as possible, we propose a network traces sampling and mixing method to generate synthetic data.
We tested our system with 24 synthetic experimental datasets that each contains 10,000 internal hosts. We implemented our P2P hosts detection component using a Map-Reduce framework, which could dramatically reduce the number of hosts subject to analysis by 99.03\% and retained all the P2P hosts in our experiments. The Map-Reduce design and implementation of our system could be deployed on popular cloud-computing platforms (e.g. amazon EC2), which ensures the scalability of our system to deal with a big data. With the best parameter settings, our system achieved 100\% detection rate with none false positives.

The rest of paper is organized as follows:
Section~\ref{sec:sec2} presents the related works.
Section~\ref{sec:sec3} explains the motivation and details of the features applied in our system.
Section~\ref{sec:sec4} describes the system design and implementation details about PeerHunter.
Section~\ref{sec:sec5} presents the experimental evaluation of PeerHunter.
Section~\ref{sec:sec7} makes the conclusion.

\section{Related Work}
\label{sec:sec2}
A few methods attempt to detect P2P botnets have been proposed \cite{gu2007bothunter, hang2013entelecheia, rahbarinia2014peerrush, zhang2014building, coskun2010friends, gu2008botminer, felix2012group, zhang2011detecting, zhang2011boosting, yen2010your, yan2013peerclean, soniya2014fuzzy, li2013gangs}. 
Host-level methods have been proposed \cite{soniya2014fuzzy}.
However, in host-level methods, all the hosts are required to be monitored individually, which is impractical in real network environments.
Network-level methods can be roughly divided into
(a) traffic signature based methods, and (b) group/community behavior based methods.


Traffic signature based methods \cite{hang2013entelecheia, rahbarinia2014peerrush, zhang2014building, gu2008botminer, zhang2011detecting, zhang2011boosting, yen2010your, felix2012group} rely on a variety of statistical traffic signatures.
For instance, Entelecheia \cite{hang2013entelecheia} uses traffic signatures
to identify a group of P2P bots in a super-flow
graph.
PeerRush \cite{rahbarinia2014peerrush} is a signature based P2P traffic categorization system, which can distinguish traffic from different P2P applications, including P2P botnet. Nevertheless, these methods suffer from botnets that have dynamic statistical traffic patterns.
Traffic size statistical features can be randomized or modified, since they are only based on the communication protocol design of a botnet. Traffic temporal statistical features can also act dynamically without jeopardizing any primary functions of a botnet.

Group or community behavior based methods \cite{yan2013peerclean, coskun2010friends} consider the behavior patterns of a group of bots within the same P2P botnet community. For instance, Coskun et al. \cite{coskun2010friends} developed a P2P botnets detection approach that start from building a mutual contact graph of the whole network, then attempt to utilize ``seeds'' (known bots) to identify the rest of bots within the same botnet. However, most of the time, it is hard to have a ``seed'' in advance.
Yan et al. \cite{yan2013peerclean} proposed a group-level behavior analysis based P2P botnets detection method.
However, they only considered to use statistical traffic features to cluster P2P hosts, which is subject to P2P botnets that have dynamic or randomized traffic patterns. Besides, their method cannot cope with unknown
P2P botnets, which is the common case in botnet detection \cite{wang2010peer}, because of relying on supervised classification methods (e.g. SVM).

\section{Background and Motivation}
\label{sec:sec3}

To demonstrate the features discussed in this section, we conducted some preliminary experiments using dataset shown in Table~\ref{table:D1} and Table~\ref{table:D2}.
Table~\ref{table:notations} shows the notations and descriptions, and Table~\ref{table:prelim} shows the measurements of features.
\vspace{-1.0em}
\begin{table}[h]
\footnotesize
\captionsetup{font=footnotesize}
\caption{Notations and Descriptions}
\label{table:notations}
\centering
\begin{tabular}{c|l}
\hline
\bfseries Notations & \bfseries Descriptions\\
\hline
MNF & the management network flows\\
\hline
AVGDD & the average \# of distinct /16 MNF dstIP prefixes\\
\hline
AVGDDR & the average destination diversity ratio\\
\hline
AVGMC & the average \# of mutual contacts between a pair of hosts\\
\hline
AVGMCR & the average mutual contact ratio\\
\hline
\end{tabular}
\vspace{-1.2em}
\end{table}


\begin{table}[h]
\footnotesize
\captionsetup{font=footnotesize}
\caption{Measurements of Features}
\label{table:prelim}
\centering
\begin{tabular}{l|r|r|r|r}
\hline
\bfseries Trace & \bfseries AVGDD & \bfseries AVGDDR  & \bfseries AVGMC & \bfseries AVGMCR \\
\hline
  eMule & 8,349 & 17.6\% & 3,380 & 3.7\%\\
  \hline
  FrostWire & 11,420 & 15.2\% & 7,134 & 4.5\%\\
  \hline
  uTorrent & 17,160 & 8.7\% & 13,888 & 3.5\%\\
  \hline
  Vuze & 12,983 & 10.1\% & 18,850 & 7.9\%\\
  \hline
  Storm & 7,760 & 25.1\% & 14,684 & 30.2\%\\
  \hline
  Waledac & 6,038 & 46.0\% & 7,099 & 37.0\%\\
  \hline
  Sality & 9,803 & 9.5\% & 72,495 & 53.2\%\\
  \hline
  Kelihos & 305 & 97.4\% & 310 & 98.2\%\\
  \hline
  ZeroAccess & 246 & 96.9\% & 254 & 100.0\%\\
\hline
\end{tabular}
\vspace{-1.5em}
\end{table}
\subsection{P2P Network Characteristics}
\label{sec:sec3_1}
Due to the decentralized nature of P2P network, a P2P host usually communicates with peers that distributed in a large range of distinct physical networks, which results in the destination diversity (DD) characteristic \cite{rahbarinia2014peerrush} of P2P management network flow (MNFs). MNF is the network flow for maintaining the function and structure of the P2P network. The P2P network flow mentioned in this section and the rest only refers to P2P MNF.



We use DD as our main feature to detect P2P network flows and further identify P2P hosts. In addition, we use the number of distinct /16 IP prefixes of each host's network flows, rather than BGP prefix used in \cite{zhang2014building} to approximate DD feature of each P2P host/network flow. /16 IP prefix is a good approximation of network boundaries. For instance, it is very likely that two IP addresses with different /16 IP prefixes belong to two distinct physical networks. This is also supported by Table~\ref{table:prelim}, which shows the network flows in a P2P network spreading across a large number of distinct physical networks according to the number of /16 IP prefixes.
\vspace{-0.3em}
\subsection{Mutual Contacts}
\label{sec:sec3_2}
\begin{figure}[tb]
\captionsetup{font=footnotesize}
        \centering
        \begin{subfigure}[b]{0.34\textwidth}
                \includegraphics[width=\textwidth]{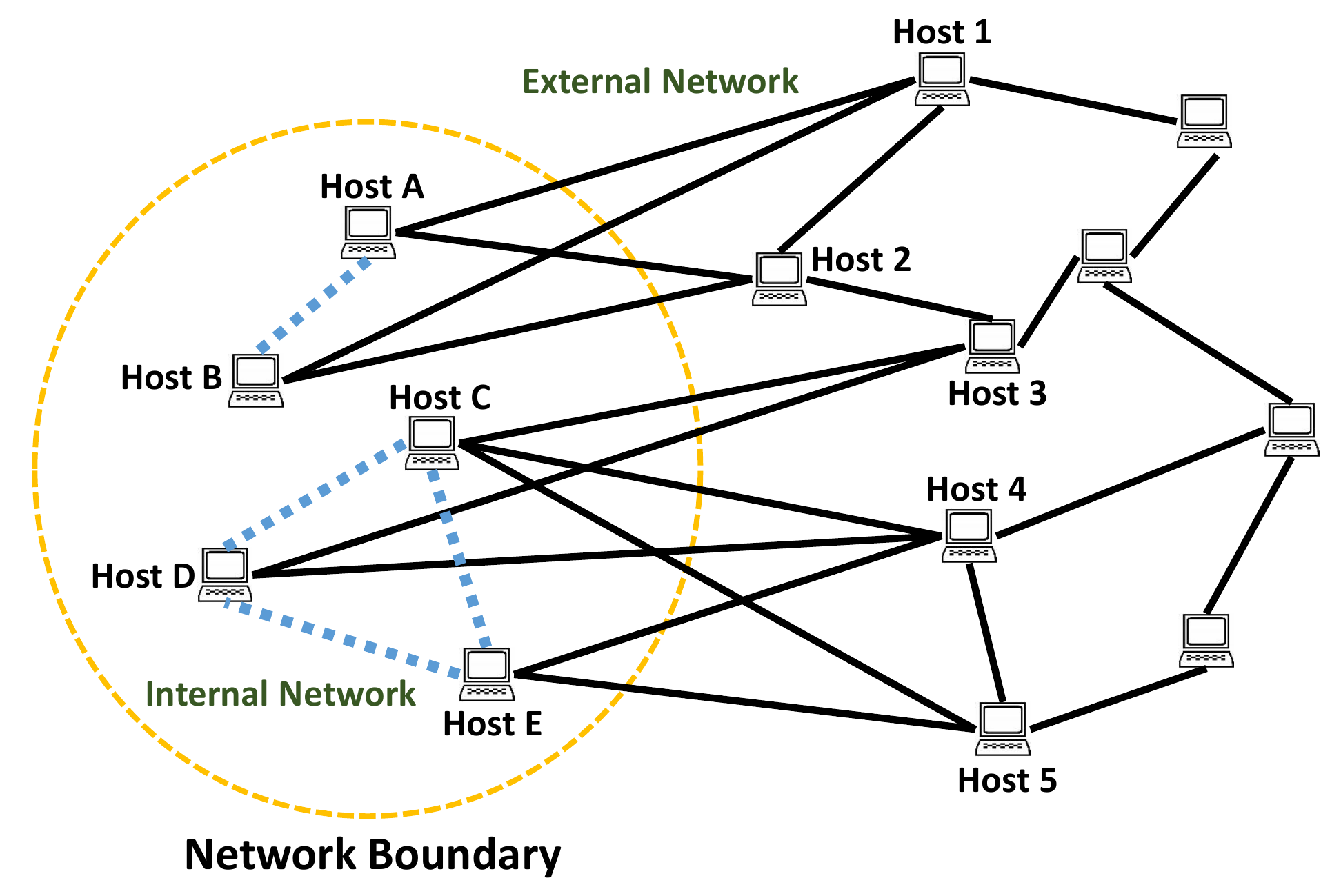}
                \caption{}
                \label{fig:network1}
        \end{subfigure}%
        ~ 
        \begin{subfigure}[b]{0.14\textwidth}
                \includegraphics[width=\textwidth]{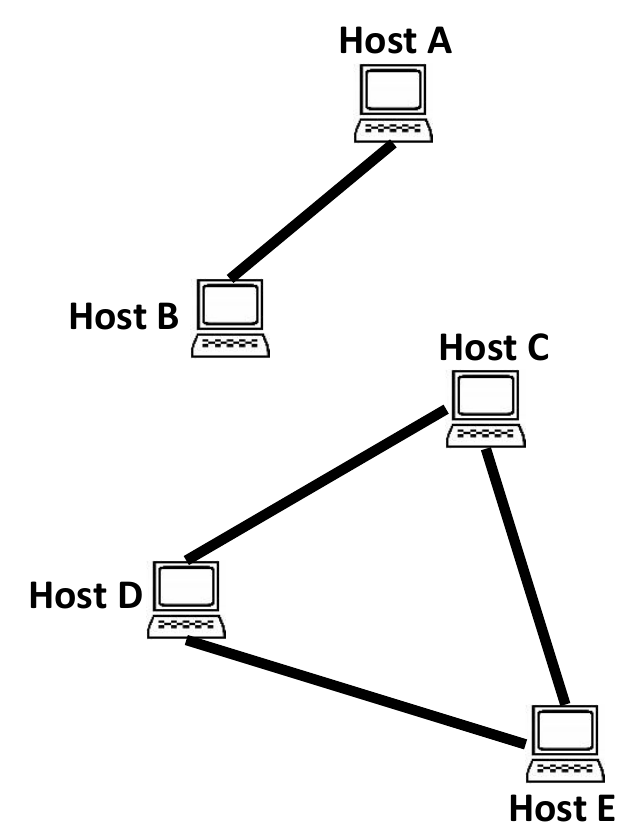}
                \caption{}
                \label{fig:network2}
        \end{subfigure}
        \caption{Illustration of network (a) and its mutual contact graph (b).}
        \label{fig:MCG}
\end{figure}

The mutual contacts (MC) between a pair of hosts is a set of shared contacts between the corresponding pair of hosts. Consider the network illustrated in Fig.~\ref{fig:network1} which contains an internal network (Host A, B, C, D and E) and an external network (Host 1, 2, 3, 4 and 5). A link between a pair of hosts means they have at least one connection. In Fig.~\ref{fig:network1}, Host 1, 2 are the mutual contacts shared by Host A, B.

Mutual contacts is the natural characteristic of P2P botnet. Compared with legitimate hosts, a pair of bots within the same P2P botnet has a much higher probability to share a mutual contact \cite{coskun2010friends}. Because bots within the same P2P botnet tend to receive or search for the same C\&C messages from the same set of botmasters (peers) \cite{holz2008measurements}. Moreover, in order to prevent peers from churning in a P2P botnet, botmaster has to check each bot periodically, which results in a convergence of contacts among peers within the same botnet \cite{zhang2014building}. However, since bots from different botnets are controlled by different botmasters, they won't share many mutual contacts. Legitimate host pairs may have a small set of mutual contacts, since nearly all hosts communicate with several extremely popular servers, such as google.com, facebook.com \cite{coskun2010friends}. Furthermore, the host pairs running the same P2P applications may also result in a decent ratio of mutual contacts, if they are accessing the same resource from the same set of peers by coincidence. However, in reality, legitimate P2P hosts with different purposes will not search for the same set of peers. Thus, we can use mutual contacts as a feature to cluster the bots within the same botnet.


The basic idea is to build a mutual contacts graph (MCG) as shown in Fig.~\ref{fig:MCG}, where Host A, B are linked together in Fig.~\ref{fig:network2}, since they have mutual contacts Host 1, 2 in Fig.~\ref{fig:network1}. Similarly, Host C, D, E are linked to each other in Fig.~\ref{fig:network2}, since every pair of them share at least one mutual contacts in Fig.~\ref{fig:network1}. More details about MCG is discussed in Section~\ref{sec:sec4_3}.

\vspace{-0.3em}
\subsection{Community Behavior Analysis}
\label{sec:sec3_3}
We consider three types of community behaviors: (a) flow statistical feature, (b) numerical community feature and (c) structural community feature.

\subsubsection{Flow Statistical Feature}
\label{sec:sec3_3_1}
Botnet detection methods using flow statistical features, have been widely discussed \cite{zhang2011detecting, saad2011detecting, livadas2006usilng, gu2008botminer}.
We use the statistical features of P2P MNFs, which are usually generated through the same P2P protocol for a specific P2P application, and some of the statistical patterns of P2P MNFs fully depend on protocols. However, the other network flows, such as data-transfer flows, are usually situation-dependant, which vary a lot even in the same P2P network. In this work, we use the ingoing and outgoing bytes-per-packets (BPP) of network flows in one P2P network as the community flow statistical feature.


\subsubsection{Numerical Community Feature}
\label{sec:sec3_3_2}
We consider two types of features: average destination diversity ratio (AVGDDR) and average mutual contacts ratio (AVGMCR).

{\bf Average Destination Diversity Ratio: }
This captures the ``P2P behavior'' of P2P botnet communities.
The destination diversity (DD) of a P2P host is the number of distinct /16 IP prefixes of each host's network flows. The destination diversity ratio (DDR) of each host is its DD divided by the total number of distinct destination IPs of its network flows.

Due to the decentralized nature of P2P networks, P2P network flows tend to have higher DDR than non-P2P network flows.
Furthermore, network flows from P2P botnet communities usually have higher average DDR (AVGDDR) than network flows from legitimate network communities. Network flows from bots within the same botnet tend to have similar DDR, since those bots are usually controlled by machines, rather than humans. However, the destinations of legitimate P2P network flows are usually user-dependant, which result in their DDR varying greatly from user to user. Besides, our botnet community detection method
aims to cluster bots within the same botnets together, rather than clustering the same legitimate P2P hosts together. Legitimate communities might contain both P2P hosts and non-P2P hosts, leading to lower AVGDDR than botnet communities.


Table~\ref{table:prelim} shows the number of distinct destination IP /16 prefixes in MNFs of each type of P2P host, where both legitimate hosts and bots spread across a large number of distinct networks. However, most of the botnets communities have higher AVGDDR than legitimate communities, except Sality. We could combine the next feature to identify Sality.

{\bf Average Mutual Contacts Ratio: }
This captures the ``botnet behavior'' of P2P botnet communities. The mutual contacts ratio (MCR) between a pair of hosts is the number of mutual contacts between them, divided by the number of total distinct contacts of them. This idea is based on three observations: (a) P2P botnet communities are usually formed by at least two bots, otherwise they cannot act as a group, (b) MCR between a pair of bots within the same botnet is much higher than that between a pair of legitimate hosts or bots from different botnets, and (c) each pair of bots within the same botnet has similar MCR. Thus, we consider the average MCR (AVGMCR) among all pairs of hosts within one network community as another numerical community feature.


Table~\ref{table:prelim} shows the average number of mutual contacts between a pair of hosts within the same community, where both botnets and certain legitimate network communities have a considerable number of mutual contacts. That is because those legitimate communities have much more contacts than botnets. However, botnets has much higher AVGMCR.

\subsubsection{Structural Community Feature}
\label{sec:sec3_3_3}
This captures the structural characteristics of a botnet.
The basic idea is that, every pair of bots within the same botnet tends to have a considerable number or ratio of mutual contacts. Therefore, if we consider each hosts as a vertex and link an edge between a pair of hosts if they have a certain amount or ratio of mutual contacts, the bots within the same botnet tend to form certain complete graphes (cliques). On the contrary, the contacts of different legitimate hosts usually tend to diverge into different physical networks. Thus, the probability that legitimate communities form certain cliques is relatively low. Then, we can consider P2P botnets detection as a clique detection problem, which detects cliques from a given network with certain requirements. However, since clique detection problem is NP-complete, we cannot just apply such method to detect botnets. Therefore, we use all three botnet community behaviors.

\begin{figure*}[tb]
\centering
\includegraphics[width=475pt]{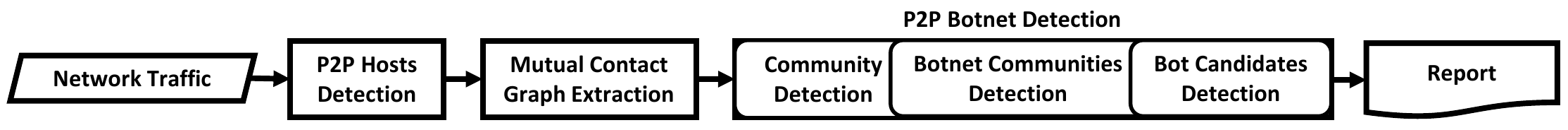}
\caption{System Overview}
\label{fig:sysview}
\vspace{-1.5em}
\end{figure*}

\section{System Design}
\label{sec:sec4}
PeerHunter has three components, that work synergistically to (a) detect P2P hosts, (b) construct mutual contact graph, and (c) detect bots.
Fig.~\ref{fig:sysview} illustrates the framework of PeerHunter.

\vspace{-0.3em}
\subsection{P2P Hosts Detection}
\label{sec:sec4_2}
This component is responsible for detecting hosts engaged in P2P communications. 
The input is a 5-tuple network flow [$ip_{src}$, $ip_{dst}$, $proto$, $bpp_{out}$, $bpp_{in}$], where $ip_{src}$ is source IP, $ip_{dst}$ is destination IP, $proto$ is tcp or udp, and $bpp_{out}$ and $bpp_{in}$ are outgoing and ingoing $BPP$ of network flows. First, we cluster all network flows $F=\{f_{1}$, $f_{2}$, \dots, $f_{k}\}$ based on the 4-tuple [$ip_{src}$, $proto$, $bpp_{out}$, $bpp_{in}$] into flow clusters $FC=\{FC_{1}$, $FC_{2}$, \dots, $FC_{m}\}$. 
Then, we calculate the number of distinct /16 prefixes of $ip_{dst}$ (destination diversity) associated with each flow cluster, $dd_{i}=DD(FC_{i})$. If $dd_{i}$ is greater than a pre-defined threshold $\Theta_{dd}$, we consider $FC_{i}$ as a P2P MNF cluster, and the corresponding source hosts as P2P hosts.

\begin{algorithm}
\caption{P2P Hosts Detection}\label{alg:alg1}
\begin{algorithmic}[1]
\Function{Map}{$[ip_{src}, ip_{dst}, proto, bpp_{out}, bpp_{in}]$}
\State $Key \gets [ip_{src}, proto, bpp_{out}, bpp_{in}]$
\State $Value \gets ip_{dst}$
\State \textbf{output} $(Key, Value)$
\EndFunction
\Function{Reduce}{$Key$, $Value[$ $]$}
\State $k \gets Key$
\State $dd_{k}=\O$
\For {$v \in Value[$ $]$}
    \State $dd_{k} \gets dd_{k} \cup \{v\}$
\EndFor
\If {$|dd_{k}| \geq \Theta_{dd}$}
\For {$v \in Value[$ $]$}
    \State \textbf{output} $(k, v)$
\EndFor
\EndIf
\EndFunction
\end{algorithmic}
\end{algorithm}
\vspace{-0.4em}
As shown in Algorithm~\ref{alg:alg1}, we design this component using a MapReduce framework \cite{dean2008mapreduce}. For a mapper, the input is a set of 5-tuple network flows, and the output is a set of key-value pairs.
For a reducer, the input is the set of key-values pairs.
Then, the reducer aggregates all values with the same key to calculate the DD of each flow cluster, and finally output the detected P2P MNFs based on $\Theta_{dd}$.
\vspace{-0.4em}
\subsection{Mutual Contact Graph Extraction}
\label{sec:sec4_3}
This component is responsible for extracting mutual contact graph (MCG) through mutual contacts. The input is a list of detected P2P hosts, $H$=$\{h_{1}$, $h_{2}$, $\ldots$, $h_{|H|}\}$
, and their corresponding P2P MNFs, $F$=$\{f_{1}^1$, $f_{2}^1$, $\ldots$, $f_{n_{1}}^1$, $f_{1}^2$, $f_{2}^2$, $\ldots$, $f_{n_{2}}^2$, $\ldots$, $f_{1}^{|H|}$, $f_{2}^{|H|}$, $\ldots$, $f_{n_{|H|}}^{|H|} \}$, where $f_{i}^j$ is flow $i$ from $h_{j}$. The output is a MCG, $G_{mc} = (V, E)$, where each vertex $v_{i} \in V$ contains a DDR score $ddr_{i}$ of $h_{i}$'s MNFs, and each edge $e_{ij} \in E$ contains a nonnegative MCR weight $mcr_{ij}$ between $h_{i}$ and $h_{j}$. Algorithm~\ref{alg:alg2} shows the main steps in this component.

First, for each host $h_{i}$, we generate a contact set $C_{i}$, that contains all the destination IPs in its MNFs. Each host $h_{i}$ also contains a flow statistical pattern set $S_{i}$, which contains all [$proto$, $bpp_{out}$, $bpp_{in}$] 3-tuple in its MNFs. Let $DD(C_{i})$ be the set of distinct /16 prefixes of all the IPs in $C_{i}$. Then, $ddr_{i}$ and $mcr_{ij}$ can be calculated as below.

\begin{equation}
    \label{eq:ddr}
    ddr_{i} = \frac{\| DD(C_{i}) \|}{\| C_{i} \|}   \ \ \ \ \ mcr_{ij} = \frac{C_{i} \cap C_{j}}{C_{i} \cup C_{j}}
\end{equation}

Furthermore, as discussed in Section~\ref{sec:sec3_3_1}, MNFs from different hosts within the same network communities should have similar statistical patterns. Thus, for each pair of input hosts, say $h_{i}$ and $h_{j}$, we calculate the intersection between $S_{i}$ and $S_{j}$. If $S_{i} \cap S_{j} = \O$, then there is no edge between $h_{i}$ and $h_{j}$ in MCG. Otherwise, they share at least one MNF statistical pattern, and we calculate $mcr_{ij}$ as shown in~(\ref{eq:ddr}). Let $\Theta_{mcr}$ be a pre-defined threshold. Then, if $mcr_{ij} > \theta_{mcr}$, there is an edge between $h_{i}$ and $h_{j}$, with weight $mcr_{ij}$. Otherwise, there is no edge between $h_{i}$ and $h_{j}$ ($mcr_{ij}=0$).

\begin{algorithm}
\caption{Mutual Contact Graph Extraction}\label{alg:alg2}
\begin{algorithmic}[1]
\renewcommand{\algorithmicrequire}{\textbf{input:}}
\renewcommand{\algorithmicensure}{\textbf{output:}}
\Require $H$, $F$, $\Theta_{mcr}$
\Ensure $G_{mc} = (V, E)$

\State $E=\O$, $V=\O$
\For {$h_{i} \in H$}
    \State   $C_{i}=\O$
    \State   $S_{i}=\O$
\EndFor

\For {$f_{i}^j \in F$}
    \State $C_{j} \gets C_{j} \cup \{ip_{dst}\}$
    \State $S_{j} \gets S_{j} \cup \{[proto, bpp_{out}, bpp_{in}]\}$
\EndFor

\For {$h_{i} \in H$}
    \State $ddr_{i} \gets \frac{\| DD(C_{i}) \|}{\| C_{i} \|}$
    \State $vertex$ $v_{i} \gets <ddr_{i}>$
    \State $V \gets V \cup \{v_{i}\}$
\EndFor

\For {$\forall$ $h_{i}, h_{j} \in H$ and $i<j$}
\If {$S_{i} \cap S_{j} \neq \O$}
    \State $mcr_{ij} \gets \frac{C_{i} \cap C_{j}}{C_{i} \cup C_{j}}$.
    \If {$mcr_{ij} > \Theta_{mcr}$}
        \State $edge$ $e_{ij} \gets <mcr_{ij}>$
        \State $E \gets E \cup \{e_{ij}\}$
    \EndIf
\EndIf
\EndFor

\State \Return $G_{mc} = (V, E)$
\end{algorithmic}
\end{algorithm}

\vspace{-0.4em}
\subsection{P2P Botnet Detection}
\label{sec:sec4_4}
This component is responsible for detecting P2P bots from given MCG. First, we cluster bots
into
communities. 
Then, we detect botnet communities using numerical community behavior analysis. In the end, we perform structural community behavior analysis to further identify or verify each bot candidates. Algorithm~\ref{alg:alg3} shows the main steps.

\subsubsection{Community Detection}
In a MCG $G_{mc} = (V, E)$, $\forall$ $e_{ij} \in E$, we have $mcr_{ij} \in [0, 1]$, where $mcr_{ij}=1$ means all contacts of $h_{i}$ and $h_{j}$ are mutual contacts and $mcr_{ij}=0$ means there is no mutual contacts between $h_{i}$ and $h_{j}$. Furthermore, bots within the same botnet tend to have a large number/ratio of mutual contacts. Then, the bots clustering problem can be considered as a classical community detection problem. Various community detection methods have been discussed in \cite{fortunato2010community}. In this work, we utilize Louvain method, a modularity-based community detection algorithm \cite{blondel2008fast}, due to (a) its definition of a good community detection result (high density of weighted edges inside communities and low density of weighted edges between communities) is perfect-suited for our P2P botnet community detection problem; (b) it outperforms many other modularity methods in terms of computation time \cite{blondel2008fast}; and (c) it can handle large network data sets (e.g. the analysis of a typical network of 2 million nodes only takes 2 minutes \cite{blondel2008fast}).

Given $G_{mc} = (V, E)$ as input, Louvain method outputs a set of communities $Com=\{com_{1}, com_{2}, \ldots, com_{|Com|}\}$, where $com_{i}=(V_{com_{i}}, E_{com_{i}})$. $V_{com_{i}}$ is a set of hosts in $com_{i}$. $E_{com_{i}}$ is a set of edges, where $\forall$ $e_{jk} \in E_{com_{i}}$, we have $e_{jk} \in E$ and $v_{j}, v_{k} \in V_{com_{i}}$.

\subsubsection{Botnet Communities Detection}
Given a set of communities $Com$, for each community $com_{i} \in Com$, we start from calculating $avgddr_{i}$ and $avgmcr_{i}$, as shown below.

\begin{equation}
    \label{eq:avgddr}
    avgddr_{i} = \frac{\sum_{v_{j} \in V_{com_{i}}} ddr_{j}}{\|V_{com_{i}}\|}
\end{equation}
\begin{equation}
    \label{eq:avgmcr}
    avgmcr_{i} = \frac{2 \times \sum_{\forall e_{jk} \in E_{com_{i}}} mcr_{jk}}{\|V_{com_{i}}\| \times (\|V_{com_{i}}\| - 1)}
\end{equation}


We define two thresholds $\Theta_{avgddr}$ and $\Theta_{avgmcr}$. Then, $\forall$ $com_{i} \in Com$, if $avgddr_{i} \geq \Theta_{avgddr}$ and $avgmcr_{i} \geq \Theta_{avgmcr}$, we consider $com_{i}$ as a botnet community.

\subsubsection{Bot Candidates Detection}
Recall from Section~\ref{sec:sec3_3_3}, the MCG of a botnet usually has a structure of one or several cliques. Therefore, we utilize a maximum clique detection method $CliqueDetection$ to further identify or verify each bot candidates from botnet communities. Each time it tries to detect one or several maximum cliques on the given MCG of botnet communities. If maximum clique (at least contains 3 vertices) has been found, we consider the hosts in that clique as bot candidates, remove those hosts from the original MCG, and run the maximum clique detection algorithm on the remaining MCG, until no more qualified maximum cliques to be found, then return the set of bot candidates.

\begin{algorithm}
\caption{P2P Botnet Detection}\label{alg:alg3}
\begin{algorithmic}[1]
\renewcommand{\algorithmicrequire}{\textbf{input:}}
\renewcommand{\algorithmicensure}{\textbf{output:}}
\Require $G_{mc}$, $\Theta_{avgddr}$, $\Theta_{avgmcr}$
\Ensure $S_{bot}$

\State $S_{botnetCom}=\O$, $S_{bot}=\O$
\State $Com \gets {\bf Louvain}(G_{mc})$

\For {$com_{i} \in Com$}
    \State $avgddr_{i} \gets \frac{\sum_{v_{j} \in V_{com_{i}}} ddr_{j}}{\|V_{com_{i}}\|}$
    \State $avgmcr_{i} \gets \frac{2 \times \sum_{\forall e_{jk} \in E_{com_{i}}} mcr_{jk}}{\|V_{com_{i}}\| \times (\|V_{com_{i}}\| - 1)}$
    \If {$avgddr_{i} \geq \Theta_{avgddr}$ and $avgmcr_{i} \geq \Theta_{avgmcr}$}
        \State $S_{botnetCom} \gets S_{botnetCom} \cup \{com_{i}\}$
    \EndIf
\EndFor

\For {$com_{i} \in S_{botnetCom}$}
    \State $S_{bot} \gets {\bf CliqueDetection}(com_{i})$
\EndFor

\State \Return $S_{bot}$

\end{algorithmic}
\end{algorithm}

\section{Experimental Evaluation}
\label{sec:sec5}

\subsection{Experiment Setup}
\label{sec:sec5_1}
\subsubsection{Experiment Environment}
The experiments are conducted on one single PC with an 8 core Intel i7-4770 Processor, 32GB RAM, 400GB SSD and 4TB HHD, and on the 64-bit Ubuntu 14.04 LTS operating system.

\subsubsection{Data Collection and Analysis Tool}
The dataset contains three categories: (a) ordinary P2P network traces, (b) P2P botnets network traces, and (c) background network traces. In practice, all the network traces could be collected at a network boundary (e.g. firewall, gateway, etc.).

\begin{table}[!t]
\footnotesize
\captionsetup{font=footnotesize}
\caption{Traces of Ordinary P2P Networks (24 hrs)}
\label{table:D1}
\centering
\begin{tabular}{l|c|r|r|c}
\hline
\bfseries Trace & \bfseries \# of hosts & \bfseries \# of flows & \bfseries \# of dstIP & \bfseries Size\\
\hline
  eMule & 16 & 4,181,845 & 725,367 & 42.1G \\
  \hline
  FrostWire & 16 & 4,479,969 & 922,000 & 11.9G \\
  \hline
  uTorrent & 14 & 10,774,924 & 2,326,626 & 57.1G \\
  \hline
  Vuze & 14 & 7,577,039 & 1,208,372 & 20.3G \\
\hline
\end{tabular}
\vspace{-0.5em}
\end{table}

\begin{table}[!t]
\footnotesize
\captionsetup{font=footnotesize}
\caption{Traces of P2P Botnets (24 hrs)}
\label{table:D2}
\centering
\begin{tabular}{l|c|r|r|c}
\hline
\bfseries Trace & \bfseries \# of bots & \bfseries \# of flows & \bfseries \# of dstIP & \bfseries Size\\
\hline
  Storm & 13 & 8,603,399 & 145,967 & 5.1G \\
  \hline
  Waledac & 3 & 1,109,508 & 29,972 & 1.1G \\
  \hline
  Sality & 5 & 5,599,440 & 177,594 & 1.5G \\
  \hline
  Kelihos & 8 & 122,182 & 944 & 343.9M \\
  \hline
  ZeroAccess & 8 & 709,299 & 277 & 75.2M \\
\hline
\end{tabular}
\vspace{-0.5em}
\end{table}

\begin{table}[!t]
\footnotesize
\captionsetup{font=footnotesize}
\caption{Traces of Background Network}
\label{table:D3}
\centering
\begin{tabular}{c|c|c|c|c}
\hline
\bfseries Date & \bfseries Dur & \bfseries \# of hosts & \bfseries \# of flows & \bfseries Size\\
\hline
  2014/12/10 & 24 hrs & 48,607,304 & 407,523,221 & 788.7G \\
\hline
\end{tabular}
\end{table}

{\bf Ordinary P2P network traces ($D_{1}$): }
We used the dataset obtained from the University of Georgia \cite{rahbarinia2014peerrush} as our ordinary P2P network traces, which collected the network traces of 4 different popular P2P applications for several weeks. There are 16 eMule hosts, 16 FrostWire hosts, 14 uTorrent hosts and 14 Vuze hosts, and we randomly selected 24 hours network traces of each host. More details about $D_{1}$ are shown in Table~\ref{table:D1}.

{\bf P2P Botnet network traces ($D_{2}$): }
Part of our botnet network traces is also from the University of Georgia dataset \cite{rahbarinia2014peerrush}, which contains 24 hours network traces of 13 hosts infected with Storm and 3 hosts infected with Waledac. We also collected 24 hours network traces of another three infamous P2P botnets, Sality, Kelihos and ZeroAccess. These network traces were collected from the hosts intentionally infected by Kelihos, ZeroAccess, and Sality binary samples obtained from \cite{malsamples}. Furthermore, all malicious activities have been blocked with the same settings as shown in \cite{rahbarinia2014peerrush}. We collected the network traces of 8 Kelihos bots, 8 ZeroAccess bots and 5 Sality bots.
More details about $D_{2}$ are shown in Table~\ref{table:D2}.

{\bf Background network traces ($D_{3}$): }
We used the dataset downloaded from the MAWI Working Group Traffic Archive \cite{mawi} as background network traces, as shown in Table~\ref{table:D3}. This dataset contains 24 hours anonymized network traces at the transit link of WIDE (150Mbps) to the upstream ISP on 2014/12/10 (sample point F). This network traces contains approximate 407,523,221 flows and 48,607,304 unique IPs. 79.3\% flows are TCP flows and the rest are UDP flows.
We utilize ARGUS \cite{argus} to process and cluster network traces into the 5-tuple format tcp/udp flows. 

\subsubsection{Experimental Dataset Generation}
To evaluate our approach, we generate 24 experimental datasets by mixing the network traces from $D_{1}$ and $D_{2}$ into different sub-datasets of $D_{3}$. Table~\ref{table:ED} illustrates the summaries of experimental datasets (EDs). Each experimental datasets contains 10,000 internal hosts sampled from $D_{3}$, where the network traces of 37 randomly selected hosts are mixed with $D_{2}$, and the network traces of another 60 randomly selected hosts are mixed with $D_{1}$. To make the experimental evaluation as unbiased and challenging as possible, below we propose
two criterions.

\begin{table}[!t]
\footnotesize
\captionsetup{font=footnotesize}
\caption{Summaries of Experimental Datasets (EDs)}
\label{table:ED}
\centering
\begin{tabular}{l||r}
\hline
\bfseries Descriptions & Values\\
\hline
the \# of EDs & 24\\
\hline
the \# of bots in each ED & 37\\
\hline
the \# of ordinary P2P hosts in each ED & 60\\
\hline
the \# of internal hosts in each ED & 10,000\\
\hline
the AVG \# of external hosts in each ED & 6,607,714\\
\hline
the AVG \# of flows in each ED & 91,240,099\\
\hline
the duration of each ED & 24 hr\\
\hline
\end{tabular}
\end{table}

{\bf Maintain a bipartite network structure.}
Our system aims to deploy at a network boundary (e.g. firewall, gateway, etc.), where the network forms a bipartite structure, and only network flows within the connections between internal hosts and external hosts could be captured. Then, the network in each experimental dataset should maintain a bipartite network structure, where any pair of internal hosts should not have any communications to each other.

{\bf Keep the connectedness of mutual contacts graph.}
The easiest way to obtain a list of background hosts is to sample the hosts randomly from $D_{3}$, with the respect of bipartite structure. However, since $D_{3}$ contains an extremely large number of hosts, simply sampling hosts randomly will result in that most of the sampled background hosts do not have a mutual contact with the other background hosts, which is much easier for PeerHunter to identify botnet communities. Because less number of mutual contacts among legitimate hosts means more disconnected legitimate communities in the corresponding MCG, which is in favor of Louvain method to detect strongly connected botnet communities. Therefore, we need to sample a list of internal hosts in a way that every internal host should have at least one mutual contact with at least one another internal host.

To follow the criterions described above without making our evaluation tasks any easier, we propose the following experimental dataset generation procedure:

$\bullet$ Utilize a two-coloring approach to sample the network traces of 10,000 background hosts from $D_{3}$ without jeopardize the bipartite network structure and the connectedness of mutual contacts graph: (a) initialize two counters, $C_{black}$ and $C_{white}$, to count the number of hosts colored in black and white respectively; (b) coloring a random host $h_{i}$ as black, and $C_{black}$ plus one; (c) coloring all contacts of $h_{i}$ as white, and increase $C_{white}$ by the number of hosts colored as white in this round; (d) for each new colored host, color its contacts with the opposite color, and adjust the counters repeatedly, until we have $C_{black} \geq 10,000$ and $C_{white} \geq 10,000$; (e) select the colored host set with exactly 10,000 hosts as the internal hosts, the hosts in the other colored host set will be the external hosts; and (f) extract the network traces of the 10,000 internal hosts from $D_{3}$. Then, it forms a bipartite graph, where each colored host set forms a bipartite component, and each host shares at least one mutual contacts with some other hosts from its own bipartite component.

$\bullet$ To maintain a bipartite network structure of botnets and ordinary P2P network traces, we eliminate all communications among bots in $D_{2}$ and legitimate P2P hosts in $D_{1}$.

$\bullet$ To mix $D_{1}$ and $D_{2}$ with $D_{3}$, each time we randomly select 97 internal hosts from one sub-datasets sampled from $D_{3}$, map those IPs to 37 bots' IP in $D_{2}$ and 60 legitimate P2P hosts' IP in $D_{1}$, and merge the corresponding network traces.

To evaluate our system, 24 synthetic experimental datasets have been created by running this procedure repeatedly.

\subsection{Evaluation on P2P Host Detection}
\label{sec:sec5_2}
We evaluate the P2P host detection with different parameter settings.
This component uses a pre-defined threshold $\theta_{dd}$ (Section~\ref{sec:sec4_2})
to detect P2P hosts. We applied this component on all 24 experimental datasets, and Table~\ref{table:EvaP2P} shows the
experimental results with different $\theta_{dd}$, ranging from 2 to 13500. If $\theta_{dd}$ is set too small, non-P2P hosts are likely to be detected as P2P hosts, which results in many false positives. For instance, when $2 \leq \theta_{dd} \leq 10$, there are, on average more than 450 non-P2P hosts have been falsely identified as P2P hosts. In contrast, if $\theta_{dd}$ is set too large, all P2P hosts will be removed, which results in false negatives. For instance, when $\theta_{dd} = 5000$, there are, on average 37 P2P hosts have been falsely discarded, and when $\theta_{dd} \geq 12000$, nearly all hosts are removed. When $20 \leq \theta_{dd} \leq 185$, it detects all P2P hosts with a very small number of false positives ($\leq 1/9903$), which demonstrates that our P2P hosts detection component is stable and effective over a large range of $\theta_{dd}$ settings. The effectiveness of $\theta_{dd}$ is also subject to the time window of the collected data. In our experiment, we used 24 hrs network traces.
The destination diversity (DD) of P2P hosts tends to grow over time. Then, $\theta_{dd}$ will be effective in a even larger range, if the time window increase.


\begin{table}[!t]
\footnotesize
\captionsetup{font=footnotesize}
\caption{Detection Rate and False Positive Rate For Different $\theta_{dd}$}
\label{table:EvaP2P}
\centering
\begin{tabular}{c|c|c||c|c|c}
\hline
\bfseries $\theta_{dd}$ & \bfseries DR & \bfseries  FP & \bfseries $\theta_{dd}$ & \bfseries DR & \bfseries FP\\
\hline
  2-10 & 97/97 & $\geq$ 450/9,903 & 500-1,000 & 81/97 & 0\\
  \hline
  15 & 97/97 & $\geq$ 8/9,903 & 5,000 & 60/97 & 0\\
  \hline
  20-25 & 97/97 & $\leq$ 1/9,903 & 10,000 & 18/97 & 0\\
  \hline
  30-185 & 97/97 & 0 & 12,500 & 5/97 & 0\\
  \hline
  200 & 89/97 & 0 & 13,500 & 0 & 0\\
\hline
\end{tabular}
\vspace{-0.5em}
\end{table}

%
%
%
%
%

\subsection{Evaluation on Community Detection}
\label{sec:sec5_3}
\begin{table}[!t]
\footnotesize
\captionsetup{font=footnotesize}
\caption{Community Detection Results For Different $\Theta_{mcr}$}
\label{table:EvaCom}
\centering
\begin{tabular}{c||c|c|c}
\hline
\bfseries $\Theta_{mcr}$ & \bfseries FLCR & \bfseries  FBCR & \bfseries FBSR\\
\hline
  0-0.25 & 0 & 0 & 0\\
  \hline
  0.5 & 0 & 0 & 2.8\\
  \hline
  1.0 & 0 & 0 & 6.4\\
\hline
\end{tabular}
\end{table}

We evaluate the performance of community detection with different parameter settings. We applied this component on the remain network flows (24 experimental datasets) after the P2P host detection (with $\theta_{dd}=50$). For each experimental dataset, this component generates a MCG $G_{mc} = (V, E)$ with a pre-defined threshold $\Theta_{mcr}$, where each edge $e_{ij} \in E$ contains a weight $mcr_{ij} \in [0, 1]$. Then, we applied Louvain method (with default resolution 1.0) on the MCG for community detection. The choice of $\Theta_{mcr}$ has an influence on the community detection results.

We evaluated the community detection performance in terms of (a) the ability to separate bots and legitimate hosts, (b) the ability to separate bots from different botnets, and (c) the ability to cluster bots within the same botnet. Let $falsely$-$clustered \ hosts$ denote the number of legitimate hosts that have been clustered with bots into the same community, $cross$-$community \ bots$ denote the number of bots of different types that have been clustered into the same community, and $split$-$communities \ botnets$ denote the number of detected communities that contain bots, subtract the number of ground truth botnets (e.g. 5 in our experiments). Then, we propose three evaluation criterions: (a) False Legitimate Cluster Rate (FLCR), which is $falsely$-$clustered \ hosts$ divided by the total number of legitimate hosts during community detection; (b) False Bot Cluster Rate (FBCR), which is $cross$-$community \ bots$ divided by the total number of bots during community detection; (c) False Botnet Split Rate (FBSR), which is $split$-$communities \ botnets$ divided by the total number of ground truth botnets.

Table~\ref{table:EvaCom} shows the results with different $\Theta_{mcr}$, ranging from 0 to 1. If $\Theta_{mcr}$ is set too small, there will be more non-zero weight edges, which might result in less but larger communities. In contrast, if $\Theta_{mcr}$ is set too large, most of the vertices will be isolated, which results in more but small communities. As shown in Table~\ref{table:EvaCom}, when $\Theta_{mcr} \leq 0.25$, FBSR also remains 0, which means no botnets have been falsely split into different communities. However, as $\Theta_{mcr}$ increasing from $0.5$ to $1$, FBSR is also increasing, which means bots within the same botnets have been clustered into different communities. This reflects that most of the MCG edge weighs between bots are less than $0.5$. If $\Theta_{mcr} \geq 0.5$, bots even within the same botnets will be isolated. FLCR and FBCR are always 0 no matter how $\Theta_{mcr}$ has been changed. FLCR is 0 means that all bots are successfully separated from legitimate hosts. FBCR is 0 means none of the communities contains more than one type of bots. This results demonstrate that our system is very effective and robust in separating bots and legitimate hosts, and separating different types of bots.

\subsection{Evaluation on Botnet Detection}
\label{sec:sec5_4}

\begin{table}[!t]
\footnotesize
\captionsetup{font=footnotesize}
\centering\caption{DR and FPR For Different $\theta_{avgddr}$ and $\theta_{avgmcr}$}
\label{table:EvaBot}
\centering
\begin{tabular}{c|c|c|c|c|c|c}
\hline
 &  & \multicolumn{5}{c}{$\theta_{avgddr}$}\\
  \hline
  $\theta_{avgmcr}$ & - & 0-0.0625 & 0.125 & 0.25 & 0.5 & 1\\
  \hline
  \multirow{2}{*}{0-0.03125}& DR & 37/37 & 32/37 & 32/37 & 16/37 & 0/37\\
   & FP & 60/60 & 32/60 & 0/60 & 0/60 & 0/60\\
  \hline
  \multirow{2}{*}{0.0625-0.25}& DR & 37/37 & 32/37 & 32/37 & 16/37 & 0/37\\
   & FP & 0/60 & 0/60 & 0/60 & 0/60 & 0/60\\
  \hline
  \multirow{2}{*}{0.5}& DR & 21/37 & 16/37 & 16/37 & 16/37 & 0/37\\
   & FP & 0/60 & 0/60 & 0/60 & 0/60 & 0/60\\
  \hline
  \multirow{2}{*}{1}& DR & 0/37 & 0/37 & 0/37 & 0/37 & 0/37\\
   & FP & 0/60 & 0/60 & 0/60 & 0/60 & 0/60\\
\hline
\end{tabular}
\end{table}

We evaluate the botnet detection component with different parameter settings. We applied this component on the remain network flows (24 experimental datasets) after previous two components (with $\theta_{dd}=50$ and $\Theta_{mcr}=0.03125$). Table~\ref{table:EvaBot} shows the
results with different $\theta_{avgddr} \in [0,1]$ and $\theta_{avgmcr} \in [0,1]$. The results support our idea that the AVGDDR of legitimate host communities is lower than most of the P2P botnets. For instance, the AVGDDR of all (60/60) legitimate host communities are less than $0.25$, but the AVGDDR of 32 out of 37 botnets are higher than $0.25$. The missing ones turned out to be 5 Sality bots, which could be detected by AVGMCR. As shown in Table~\ref{table:EvaBot}, legitimate P2P hosts have lower AVGMCR than P2P bots (e.g. $\theta_{avgmcr}=0.0625$). This experimental results demonstrate that our botnet detection component is effective (detection rate equals to 100 \% with zero false positives) and stable over a large range of $\theta_{avgddr}$ (e.g. $[0,0.0625]$) and $\theta_{avgmcr}$ (e.g. $[0.0625,0.25]$).

\begin{table*}[!t]
\footnotesize
\captionsetup{font=footnotesize}
\caption{Number of hosts identified by each component}
\label{table:EvaNum}
\centering
\begin{tabular}{c|c|c|c|c}
\hline
 - & \bfseries Before P2P detection & \bfseries After P2P detection & \bfseries After Community detection & \bfseries After Bot detection\\
\hline
  \bfseries \# of hosts & 10,000 & 97 & 97 & 37\\
\hline
\end{tabular}
\vspace{-0.5em}
\end{table*}

\begin{table*}[!t]
\footnotesize
\captionsetup{font=footnotesize}
\caption{PeerHunter Execution Time}
\label{table:EvaTime}
\centering
\begin{tabular}{c|c|c|c|c|c}
\hline
 - & \bfseries P2P Host Detection & \bfseries MCG Extraction & \bfseries Community Detection & \bfseries Bot Detection & \bfseries Total \\
\hline
 \bfseries Processing Time  & 15 minutes & 5 minutes & 18 milliseconds & 11 milliseconds & 20 minutes\\
\hline
\end{tabular}
\vspace{-1.5em}
\end{table*}

\subsection{Evaluation on PeerHunter}
\label{sec:sec5_5}
We evaluate our system according to effectiveness and scalability. Effectiveness is to evaluate the capability of our systems to detect P2P botnets, and scalability is to evaluate the practicality of our systems to deal with the real world big data. We applied PeerHunter on 24 experimental synthetic datasets, with $\theta_{dd}$=$50$, $\Theta_{mcr}$=$0.03125$, $\theta_{avgddr}$=$0.0625$ and $\theta_{avgmcr}$=$0.25$, and all results are averaged over 24 datasets. 

We use detection rate and false positive rate to measure the effectiveness. As shown in Table~\ref{table:EvaNum}, our system identified all 97 P2P hosts from 10,000 hosts, and detected all 37 bots from those 97 P2P hosts, with zero false positives. It is clear that PeerHunter is effective and accurate in detecting P2P botnets.

Our system has a scalable design based on efficient detection algorithm and distributed/parallelized computation.
Out of three components in our system, the P2P botnet detection component (community detection and botnet detection as shown in Table~\ref{table:EvaNum}) has a negligible processing time compared with the other two components. This is due to previous two components are designed to reduce a huge amount of the hosts subject to analysis (e.g. 99.03\% in our experiments).
The P2P host detection component has linear time complexity, since it scans all the input flows only once to compute the flow clusters and further identify P2P flows. However, since it is the very first component to process the input data, which could be large, it still costs the highest processing time (as shown in Table~\ref{table:EvaNum}).
To accommodate the growth of a real world input data (big data), we designed and implemented the P2P host detection component using a Map-Reduce framework, which could be deployed in distributed fashion on scalable cloud-computing platforms (e.g. amazon EC2).
The MCG extraction component requires pairwise comparison to calculate edges weights.
Let $n$ be the number of hosts subject to analysis and $m$ be the maximum number of distinct contacts of a host.
We implemented the comparison between each pair of hosts parallelly to handle the growth of $n$. If we denote $k$ as the number of threads running parallelly, the time complexity of MCG extraction is $O(\frac{n^{2}m}{k})$. 
For a given ISP network, $m$ grows over time. Since our system uses a fixed time window (24 hours), for a given ISP network, $m$ tends to be stable and would not cause a scalability issue. Besides, since the percentage of P2P hosts of an ISP network is relatively small (e.g. $3\%$ \cite{zhang2014building}), and an ISP network usually has less than 65,536 (/16 subnet) hosts, $n$ would be negligible compared with $m$. Furthermore, even if $n$ and $m$ are both big numbers, our system could use an as large as possible $k$ to adapt the scale of $n$ and $m$. In a nutshell, PeerHunter is scalable to handle the real world big data.



\section{Conclusion}
\label{sec:sec7}
In this work, we present a novel
community behavior analysis based P2P botnet detection system, PeerHunter, which operates under several challenges: (a) botnets are in their waiting stage; (b) the C\&C channel has been encrypted; (c) no bot-blacklist or ``seeds'' are available; (d) none statistical traffic patterns known in advance; and (e) do not require to monitor individual host.
We propose three types of community behaviors
that can be utilized to detect P2P botnets effectively. In the experimental evaluation, we propose a network traces sampling and mixing method to make the experiments as unbiased and challenging as possible. Experiments and analysis have been conducted to show the effectiveness and scalability of our system.
With the best parameter settings, our system can achieved 100\% detection rate with none false positives.


%
%



\bibliographystyle{IEEEtran}

\bibliography{IEEEabrv,compsac16}
%
%
%

\end{document}